# GLEAN: Generative Learning for Eliminating Adversarial Noise

**Justin Lyu Kim, Kyoungwan Woo**

[1] California Academy of Mathematics and Science
[2] Massachusetts Institute of Technology

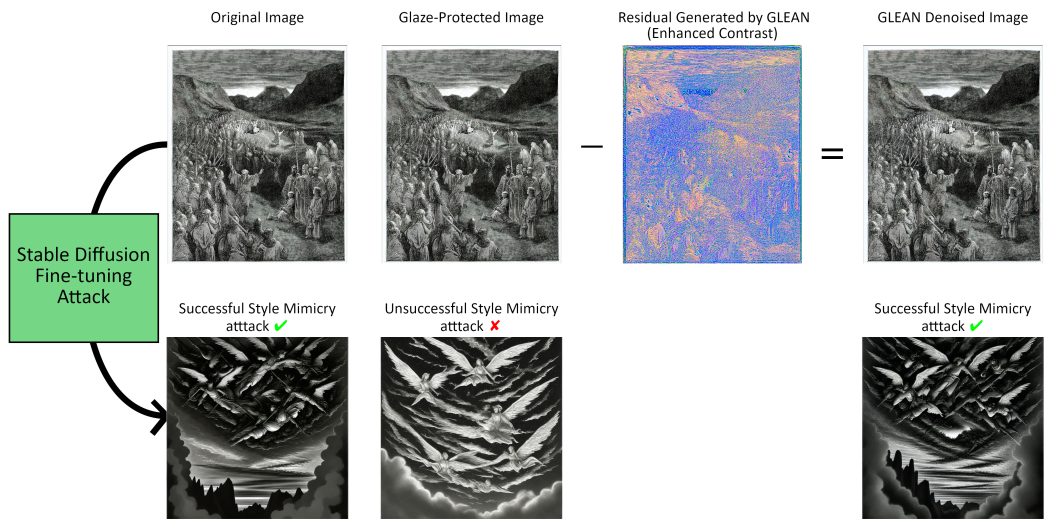

Figure 1: Diagram of GLEAN methodology above, in order of Unprotected, Glazed, GLEANed images

## Abstract

*In the age of powerful diffusion models such as DALL-E and Stable Diffusion, many in the digital art community have suffered style mimicry attacks due to fine-tuning these models on their works. The ability to mimic an artist's style via text-to-image diffusion models raises serious ethical issues, especially without explicit consent. Glaze, a tool that applies various ranges of perturbations to digital art, has shown significant success in preventing style mimicry attacks, at the cost of artifacts ranging from imperceptible noise to severe quality degradation. The release of Glaze has sparked further discussions regarding the effectiveness of similar protection methods. In this paper, we propose GLEAN- applying I2I generative networks to strip perturbations from Glazed images, evaluating the performance of style mimicry attacks before and after GLEAN on the results of Glaze. GLEAN aims to support and enhance Glaze by highlighting its limitations and encouraging further development.*

## 1  Introduction

The advent of readily-available, pre-trained generative models like Stable Diffusion [1] has lead to a rise in style mimicry attacks; using several sample artwork images, a model can be trained to generate works that plagiarize an author's original style. [2] This is often achieved by using an image-to-text model such as CLIP [3] to generate captions describing each sample artwork, and using the image-label pairs to "fine-tune" a pre-trained model to generate new works with similar styles with text guidance. This fine-tuning process is widely available; the process can leverage consumer-grade GPUs, is easily accessed through web UIs, and does not require many samples of artwork to perform. [4] This means malicious actors can leverage widely available tools to easily mimic an artist's work without consent. [2]

In response to style mimicry attacks, many tools involving image perturbation, poisoning, noising, etc have been released to the public. Specifically, a "style cloaking" tool named Glaze [5] has become popular within the digital art community. Glaze leverages Image-to-Image Translation (I2I) [6], translating an input image to an output image with various alterations. Conventional I2I tasks may involve the goal of preserving characteristics of the input image while changing other attributes, such as the resolution or objects present in the image. Similarly, Glaze translates an artwork of an original visible style A into an image that appears to models as target style B by adding a limited number of perturbations. By adding perturbations that specifically target features corresponding to the author's style, Glaze is able to "cover up" the original style and replace them with the target style B. By using this approach to "budget" how many perturbations it places on the image, Glaze attempts to minimize the visual difference its perturbations create. [5]

Several approaches have attempted to circumvent Glaze, such as introducing Gaussian noise, removing information from Glazed images, or removing perturbations from Glazed images directly. [7] For example, IMPRESS is a framework built upon the observation that Glazed images, when reconstructed using a latent diffusion model (LDM), can display the effects of its protective perturbations. [8] This provides an avenue for the latent diffusion model to train on the inconsistency loss between the original image and the LDM reconstruction. However, the effectiveness of IMPRESS is under contention by the authors of Glaze, as results have not been fully replicable. [9]

In this paper, we introduce GLEAN, a framework that utilizes an I2I Generative Adversarial Network (GAN) using Fast Fourier Convolutions (FFCs) to target the patterns associated with Glaze's perturbations. [10] [11][12] In our GAN, we reconfigure the original encoder-decoder stack of pix2pix and the structure of FFCs from the super-resolution upscaler FREDSR. [13] GLEAN is based upon the observation that Glaze creates perturbations based on the original artwork. As such, a model trained to generate Glaze's perturbations given a Glazed image would be able to subtract these perturbations and remove them from the image. This approach has been used in denoising and upscaling tasks in the past, in models such as VDSR and FREDSR. [14] To employ this approach, we train our model on the residual image between an original artwork and its Glazed counterpart to specifically target the perturbations created by Glaze.

Our work demonstrates the GLEAN framework/model: a GAN model employing FFC blocks trained to generate applied adversarial noise in various "poisoned" artworks. In testing, according to our quantitative metrics, GLEAN showed promising results in removing the perturbations generated by Glaze and making sample artworks vulnerable to style transfer attacks with minimal loss in quality. Based on our results, we attempt to provide a better understanding of the nature of models like Glaze and provide feedback that will help in strengthening protection methods.

## 2 Method

In this section we discuss the vulnerabilities and observations that our model is based around, and propose GLEAN, a framework for removing adversarial noise from artworks processed by the tool Glaze.

### 2.1 Observations and Vulnerabilities in Glaze

Glaze distinguishes itself from existing image cloaking methods, which indiscriminately apply a cloak to the entire image, by focusing on identifying style-specific features of an original artwork and utilizing them to compute style cloaks to specifically modify aforementioned features of the original artwork. [5]

Glaze initially calculates the original style of the work by utilizing a feature extractor. [5] It then computes a target style moderately different from the original style. Next, using a style transfer model, it generates a transformed version of the image in the target style. Lastly, the final style cloak is computed by transferring traits of the transformed image associated with its feature representation- traits that a machine learning model would identify as the target style. By attempting to extract only the representative traits of an art style, Glaze seeks to minimize visual perturbations to the human eye. Glaze's perturbations are most visible to humans in areas of an artwork with flat, similar colors, where they take the appearance of "ripples." We conjecture that these patterns may be the result of Glaze extracting features representative of artworks such as impasto oil paintings, which due to their use of thick, layered paint, result in similar "ripples" at the ends of each application of paint.

As the Glaze team states: this approach leads to cloaks that are dependent both on the original image's content and feature representation. [5] This is advantageous for generalizeability as cloaks can be optimized on a per-artwork basis. However, the approach causes a key vulnerability: as the style cloak is generated and optimized per image, it is possible for a machine learning model to reverse-engineer the style cloak given a cloaked image.

### 2.2 Introducing GLEAN

With this in mind, we propose our framework GLEAN: Generative Learning for Eliminating Adversarial Noise. GLEAN is a GAN model with architecture based off of a super-resolution upscaler: FREDSR. [13] FREDSR was selected in part due to its advantages in low parameter size of just 37000 and its usage of residual learning- taking advantage of bicubic upscaling and residual images. Notably, FREDSR does not modify the input image resolution, similarly to GLEAN. [13] In training, GLEAN is given a Glazed artwork with the residual label -original image subtracted from Glazed image- and attempts to generate the residual label from the Glazed artwork. Then, the generated cloak is subtracted from the Glazed image, resulting in the final GLEANed image. As GLEAN is using a GAN architecture, we pass the residual label and the output generated cloak into a simple discriminator.

GLEAN, like FREDSR, utilizes Fast Fourier Convolutions, which splits the image information into two channels: one local, calculated using standard convolutions, and one global, calculated using a 2-dimensional fast Fourier transform. [13] We hypothesized that FFC blocks will perform well in removing the perturbations created by Glaze, which often have "regular" patterns, such as the aforementioned "ripple" appearances in its cloaks.

We note that GLEAN does not require the usage of the aforementioned FREDSR model. Computational constraints and the need to run on consumer GPUs necessitated the use of a low parameter model, but larger parameter models could be used to generate more accurate style cloaks.

### 2.3 Model Architecture

See Figure 2 for GLEAN Model architecture.

## 3 Experiments

We discuss the experiments we conducted on GLEAN to assess its performance in removing Glaze's adversarial noise. We wished to utilize the CLIP based metrics Glaze utilized in the original paper- However, as of October 2023, the Glaze authors have stated that CLIP metrics perform poorly on assessing attacks on Glaze. [9] As such, we conducted a human survey to determine the style accuracy of mimicry attacks.

### 3.1 Training

GLEAN was trained on 100,000 artworks randomly sampled from a dataset with 120,000 artworks generated using

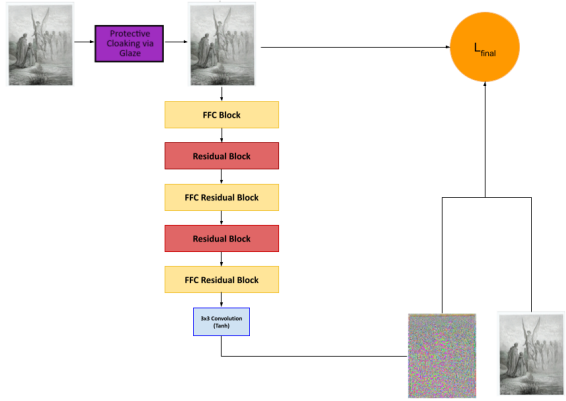

Figure 2: Model Architecture: We utilize the optimizers, learning rate decay, losses, and training methods employed by FREDSR. [13]

Stable Diffusion with varying levels of Glaze protection, with varying versions of Glaze. The training images consisted of a single resolution- 512x512. Training validation consisted of the remaining 20,000 artworks.

We then take a subset of 10 historical artworks from [15] with the same author, similar styles, and identical genres to measure metrics and evaluate the performance of GLEAN.

SSIM and PSNR during Validation

| Epoch | SSIM ↑ | PSNR↑ |
|---|---|---|
| Original/Glazed | 0.866 | 30.75 |
| 10 | 0.848 | 30.43 |
| 20 | 0.878 | 33.68 |
| 30 | 0.894 | 32.94 |
| 40 | 0.905 | 35.47 |
| 50 | 0.9132 | 36.348 |

**Table 1:** GLEAN's results show that GLEAN's output images are much closer to the original image compared to Glaze. We note that due to varying intensity in Glaze's protections, many Glazed images had heavy visual perturbations.

SSIM and PSNR for Historical Art [15]

| Artist/Style | SSIM ↑ | PSNR↑ |
|---|---|---|
| Dore/ Romanticism | 0.935 | 34.71 |
| Dore/ GLAZED | 0.855 | 31.56 |
| Rembrandt/ Baroque | 0.963 | 37.43 |
| Rembrandt/ GLAZED | 0.924 | 35.43 |
| Van Gogh/ Realism | 0.911 | 33.98 |
| Van Gogh/ GLAZED | 0.901 | 32.65 |

**Table 2:** We see measurable gains across all metrics from GLEANed images from Glazed images, showing that GLEAN is able to reduce the perturbations of Glaze by a significant amount.

### 3.2 Evaluation

After training, we process several artworks of a specific artist/style through GLEAN. We then fine-tune a pre-trained Stable Diffusion model on the original images, Glazed images, and GLEANed images, while using a pre-trained CLIP model to generate captions for each- similar to the methods described in [7]. We then attempt style mimicry using our fine-tuned Stable Diffusion model, and compare the effectiveness of style transfer in the generated images. [7]

### 3.3 Results

While there is an immediate visual difference in the style mimicry results of GLEAN versus Glaze, we compiled feedback from 20 volunteers to evaluate our results. As seen in the below table, while style mimicry attempts from Glazed images did not perform as well as the baseline or Gleaned images, Gleaned images and Original images performed similarly. Overall, our results show that GLEAN is very effective at removing style cloaks generated by Glaze.

Preference for Style Mimicry

| Artist/Style | Original | Glazed | Gleaned |
|---|---|---|---|
| Dore/ Romanticism | 0.44 | 0.20 | 0.36 |
| Rembrandt/ Baroque | 0.41 | 0.16 | 0.43 |
| Van Gogh/ Realism | 0.51 | 0.09 | 0.40 |

**Table 3:** Style mimicry performed with 5 prompts, text based guidance. Measured by 20 participants, given mimicry atttempts over all 3 categories

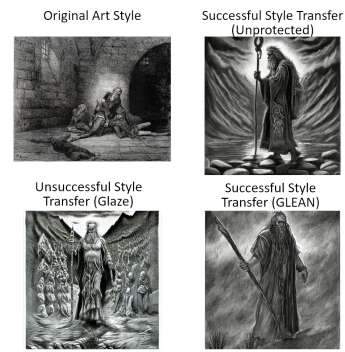

Figure 3: Example of style mimicry attack attempts

## 4 Related Work

### 4.1 Image protection approaches

The authors of Glaze have also developed Nightshade, a tool that "poisons" images, rendering them disruptive to large-scale diffusion models and image2image translation models. [16] Unlike Glaze, Nightshade's goal is to "confuse" diffusion models by generating text-image pairs that appear consistent, but with perturbed images that are significantly different from the corresponding text label in a model's latent feature space. Mist is an earlier example of image poisoning for diffusion models than Glaze, and is applied to a similar use case of preventing style transfer attacks on artists' works. [17] Additionally, works such as Anti-Dreambooth were designed to prevent fake personalized images, but showed promise in preventing style mimicry. [7][18]

### 4.2 Other methods to remove perturbations

Various other researchers have proposed methods to denoise or render ineffective adversarial perturbations generated by protection methods such as Glaze. IMPRESS is a framework that addresses adversarial noise by generating an image that is visually similar to the protected image and remains consistent when reconstructed by a latent diffusion model (LDM). [8] This is in contrast to GLEAN, which attempts to isolate and remove the perturbations directly without processing the image through a LDM. The authors of Glaze argue that IMPRESS causes quality loss in purified artworks and show weaker performance on non-historical art styles or styles such as realism. [9] Despite these limitations, by user survey conducted in [7], IMPRESS-cleaned images still performed slightly better than Glaze in terms of style transfer.

DIFFPURE, a diffusion model used for purifying adversarial noise, also showed promising results on Glaze. [7] DIFFPURE uses an img2img approach with the goal of generating a non-cloaked version of a cloaked image, similar to GLEAN. [19] However, unlike GLEAN, DIFFPURE utilizes a diffusion model rather than a GAN and generates the purified images directly instead of generating style cloaks like GLEAN. Lastly, a noisy upscaler proved effective in removing Glaze generated perturbations via addition of Gaussian noise to images before tuned Stable Diffusion upscaling. [7] The Glaze team released statements regarding the noisy upscaling and DIFFPURE attack vectors, alongside a 2.1 update for Glaze with a claimed greater resistance against attacks involving noise/information loss. The Glaze team also reported quality loss in images due to information loss in these two approaches. [20] GLEAN was trained on the aforementioned Glaze 2.1 version and show promising results.

## 5 Discussion

**5.1 Limitations** Our initial training dataset of stable-diffusion generated images presents a significant limitation, as images generated by a diffusion model may have artifacts not found in human-created artworks. We would ideally train this model on a dataset of human artworks, such as WikiArt. However, as Glaze is unable to be ran on command line interfaces and only work on macOS/Windows, we were faced with a significant lack of computing resources. Generating a Glazed dataset from larger WikiArt images would take between 4-7 months on a single NVIDIA 3090, and running multiple instances of Glaze would require a multi-GPU machine running Windows. To verify our results, evaluation was performed on historical artworks, not AI generated images. We inquired the Glaze authors on the existence of a cloud-computing compatible version of Glaze or a pre-generated dataset for training Glaze, but we did not receive a response as of this time.

Additionally, in some cases GLEAN includes stylistic features found in the original artwork (e.g. wrinkles, patterns from paint) as part of its generated Glaze cloak. This behavior is concerning as it means images processed using GLEAN may have features of the original art style removed. So far, style mimicry has been successful in these cases, but more rigorous testing may be required to determine if this is a weakness of our model. Running GLEAN on unprotected images show a very minimal detection by GLEAN, leading to minimal changes. However, it is not clear if performing GLEAN on unprotected images will lead to low style mimicry performance. As Glaze protected images are not labelled as Glaze in the wild, it would be best to perform a "best of two" approach, one utilizing GLEAN and the other performing mimicry on the unprotected image or develop a model that detects if an image has been Glazed or not.

Lastly, we do not have evidence for or against GLEAN's generalizability to removing other image protection cloaks. We believe that training GLEAN on images protected by methods other than Glaze, such as Photoguard [21] may provide further insights.

**5.2 Ethics** As Glaze is a security measure to assist artists, a tool built to "break" Glaze raises serious ethical questions. We have reached out to the Glaze team regarding GLEAN on multiple occasions, but have received no response. As such, the codebase of GLEAN will not be published until responses from the Glaze team are received. All art used in GLEAN comes from historical artists or Stable Diffusion.

## Acknowledgments

We give a huge amount of thanks to the MIT SuperCloud and CSAIL HPC clusters for providing computing resources vital to research results reported within this paper, Tensorflow as the platform of choice for our models [22], the volunteers who helped evaluate the performance of GLEAN and indicated their preferences, and various libraries mentioned in [23][24].